\documentclass[final,twocolumn,5p,times]{elsarticle}

\usepackage{hyperref}

\usepackage{adjustbox}

\journal{Journal of \LaTeX\ Templates}

\begin{document}

\begin{frontmatter}

\title{Enriched TeO$_2$ bolometers with active particle discrimination: towards the CUPID experiment}

\author[LNGS,USC]{D.R.~Artusa}
\author[USC]{F.T.~Avignone III}
\author[LBL]{J.W.~Beeman}
\author[INFN-Rm1]{I.~Dafinei}
\author[CSNSM]{L.~Dumoulin}
\author[SICCAS]{Z.~Ge}
\author[CSNSM,INSU]{A.~Giuliani}
\author[MIB]{C.~Gotti}
\author[CSNSM]{P.~de~Marcillac}
\author[CSNSM]{S.~Marnieros}
\author[GSSI,LNGS]{S.~Nagorny}
\author[LNGS]{S.~Nisi}
\author[CEA]{C.~Nones}
\author[UCB]{E.B.~Norman}
\author[CSNSM]{V.~Novati}
\author[CSNSM]{E.~Olivieri}
\author[LNGS]{D.~Orlandi}
\author[GSSI,LNGS]{L.~Pagnanini}
\author[LNGS]{L.~Pattavina}
\author[MIB]{G.~Pessina}
\author[LNGS]{S.~Pirro\corref{corresponding-author}}
\author[CSNSM,INR]{D.V.~Poda}
\author[MIB]{C.~Rusconi}
\author[GSSI,LNGS]{K.~Sch\"affner}
\author[LLNL]{N.D.~Scielzo}
\author[SICCAS]{Y.~Zhu}

\cortext[corresponding-author]{Corresponding author}

\address[LNGS]{INFN - Laboratori Nazionali del Gran Sasso, I-67010, Assergi, Italy}
\address[USC]{Dept. of Physics and Astronomy, University of South Carolina, SC 29208, Columbia, USA}
\address[LBL]{Materials Science Division, Lawrence Berkeley National Laboratory, CA 94720, Berkeley,   USA}
\address[INFN-Rm1]{INFN - Sezione di Roma, I-00185, Roma,   Italy}
\address[CSNSM]{CSNSM, Univ. Paris-Sud, CNRS/IN2P3, Universit\'e Paris-Saclay, 91405, Orsay,  France}
\address[SICCAS]{Shanghai Institute of Ceramics - Chinese Academy of Sciences, Jiading district, 201800, Shanghai, PR of China}
\address[INSU]{DISAT, Universit\`a dell'Insubria, I-22100 Como, Italy}
\address[MIB]{INFN - Sezione di Milano Bicocca, I-20126, Milano, Italy}
\address[GSSI]{INFN - Gran Sasso Science Institute, I-67100, L'Aquila, Italy}
\address[CEA]{CEA Saclay, DSM/IRFU, 91191 Gif-sur-Yvette Cedex, France}
\address[UCB]{ Dept. of Nuclear Engineering,  University  of California CA 94720, Berkeley,  USA}
\address[INR]{Institute for Nuclear Research, MSP 03680, Kyiv, Ukraine}
\address[LLNL]{Lawrence Livermore National Laboratory - Nuclear and Chemical Sciences Division, CA 94550, Livermore,  USA}

\begin{abstract}

We present the performances of two 92\% enriched  $^{130}$TeO$_2$  crystals operated as thermal bolometers in view of a 
next generation experiment to search for  neutrinoless double beta decay of $^{130}$Te.
The crystals, 435 g each, show an energy resolution, evaluated at the 2615 keV $\gamma$-line of $^{208}$Tl, of  6.5 and 4.3 keV FWHM.
The only observable  internal radioactive contamination arises from $^{238}$U (15  and 8  $\mu$Bq/kg, respectively).
The internal activity of the most problematic nuclei for neutrinoless double beta decay, $^{226}$Ra and $^{228}$Th, are both evaluated  
as $<$3.1  $\mu$Bq/kg for one crystal and $<$2.3  $\mu$Bq/kg for the second.
Thanks to the readout of the weak Cherenkov light emitted by $\beta/\gamma$ particles by means of Neganov-Luke bolometric light detectors we 
were able to perform an event-by-event identification  of $\beta/\gamma$  events with a  95\% acceptance level, while establishing a rejection factor  
of  98.21~\% and 99.99~\% for $\alpha$ particles.
  
\end{abstract}


\end{frontmatter}


\section{Introduction} 
The following three important questions in neutrino physics will be addressed by next generation neutrinoless  double beta decay (0$\nu$-DBD) experiments: are neutrinos Majorana particles that differ from antineutrinos only by helicity? Is lepton number conservation violated? What is the neutrino mass-scale? Searches for  0$\nu$-DBD have been carried out for many decades investigating a large
variety of nuclei with many different experimental techniques~\cite{Avignone-2008:481}.
However the discovery of the   atmospheric neutrinos oscillations by Super-Kamiokande as well as those observed in  solar neutrinos by the  SNO experiment - both awarded  the Nobel Prize in 2015 - boosted these searches, and  now is  an optimum time to launch next generation 0$\nu$-DBD experiments. 
Recent analyses of all of the  atmospheric, solar, and reactor neutrino oscillations~\cite{Capozzi-2016:218}   indicate that there exist scenarios in which the effective Majorana mass of the electron neutrino could be larger than 0.05 eV. 
Within the last few years, the most stringent limits on 0$\nu$-DBD  came from EXO-200~\cite{Albert-2014:229}, GERDA~\cite{Agostini-2016:1876} and CUORE-0~\cite{Alfonso-2015:102502} while, very recently,  the KamLAND-Zen experiment~\cite{Gando-2016:082503} set the strongest  limit on this   decay,   using $^{136}$Xe.
While recent experimental   achievements  are impressive, it is difficult to compare results from different isotopes because of the large uncertainties in the nuclear matrix elements. 
Ultimately, the  goal of  next the generation 0$\nu$-DBD experiments  is to sensitively probe the entire inverted hierarchy region. 
To reach this sensitivity, the total masses of parent isotopes must be increased using enriched isotopes, and the backgrounds drastically decreased.
A ton-scale $^{76}$Ge  experiment (with a possible common effort between GERDA and MAJORANA)~\cite{Abgrall-2014:365432},  the full EXO experiment (nEXO~\cite{Albert-2015:015503}), and a possible upgrade of the CUORE experiment (CUPID, CUORE Upgrade with Particle IDentification~\cite{CUPID-2015,Artusa2014:3096}) are all designed to achieve this goal.

The advantage of the bolometric technique proposed for CUPID is not only the possibility to choose different DBD emitters combined with the capability of having a high resolution detector,  
but the realization of double-readout detectors in order to perform an active  particle discrimination to reject the natural background.
CUPID is presently  in an R\&D~ phase testing  different type of crystals containing  most of the interesting  DBD emitters ($^{82}$Se, $^{100}$Mo, $^{116}$Cd, $^{130}$Te).
The aim of CUPID, which  will use the CUORE infrastructure once that experiment has concluded, is to increase  the sensitivity   to completely cover the inverted hierarchy region. In order to reach this goal, two major scientific milestones need to be reached:
\begin{enumerate}
	\item increase the number of active DBD nuclei through development of bolometers made of   
	  enriched isotopes (as the experimental volume of the CUORE cryostat is fixed);
	\item decrease the present natural radioactive background by two orders of magnitude by rejecting  the major source  of background for DBD bolometers due to $\alpha$-particle interactions~\cite{Alessandria-2013:13}.
\end{enumerate}

The initial idea to decrease the $\alpha$-background in DBD bolometers was to use scintillating crystals~\cite{Pirro-2006:2109}  in which the discrimination 
between e/$\gamma$ and $\alpha$/neutron  particles can be simply obtained with  the additional readout of the scintillation  light, through a second - very sensitive - bolometer 
working as a Light Detector (LD).
Rather recently, however,  after the observation of a very tiny  light signal  in a small TeO$_2$ bolometer~\cite{Coron-2004:159},  it was suggested~\cite{Tabarelli-2010:359} 
that particle discrimination could be obtained also in non-scintillating crystal bolometers  (like TeO$_2$) by exploiting the Cherenkov light emission. 
Heavy $\alpha$ particles arising from natural radioactivity  have velocities far below the threshold to emit Cherenkov photons in any kind of crystal. 
In the last four years several  tests were performed on large~\cite{Beeman-2012:558,Pattavina-2016:186,Schaeffner-2015} and very small~\cite{Willers:2015,Gironi-2016}  TeO$_2$  crystal samples coupled with different types of bolometric LDs.  
The challenge of this method  is the detection of the  extremely small amount of light emitted by electrons at the   0$\nu$-DBD energy of $^{130}$Te  (2.53 MeV)  that is of the order of $\approx$100 eV~\cite{Casali-2015:12}. 

In this work we present for the first time the performance of \textit{large enriched} TeO$_2$ in which the Cherenkov light is used for particle identification.
This work  demonstrates that   $^{130}$TeO$_2$ can be a suitable candidate for the CUPID experiment in terms of energy resolution, internal radioactive contaminations and $\alpha$-background discrimination.

\section{Enriched crystal growth}
The  $^{130}$TeO$_2$ crystals used in this work were manufactured starting from enriched  $^{130}$Te  in the form of metal powder, purchased from JSC Isotope, Russia. 

The purity of the enriched material was certified  as $>$~99.9875~\%. The concentrations of   the most troublesome  metallic impurities were measured  independently by ICP-MS and were found to be below 1 ppm, except for Fe (1.5 ppm), Cu (3.5 ppm) and Al (4.5 ppm).  Radioactive $^{238}$U and $^{232}$Th were not observed, with a 
detection limit of the order of 5 ppt.
The isotopic abundances of Tellurium in the powder, as measured by ICP-MS, are given in 
Table~\ref{tab:isotopic-concentration} and compared to the values  reported by the vendor and the isotopic concentration of natural Tellurium~\cite{Meija-2016:265}.
\begin{table}[t] 
\centering
\caption{Concentration of the most abundant  Tellurium isotopes in the metal used for the production of the  crystals in this work. The errors on the measurements are
of the order of  0.5~\% for the first two rows, and of the order of 10~\% for the other three.}
\label{tab:isotopic-concentration}       
\begin{tabular}{lcccc}
\hline
Isotope    					&ICP-MS		 			      				&Certification 								      &Natural 			             					 		   	\\
													  & [\%] 		 						            &[\%]       				   		  				&[\%]    																 	     \\
\hline
$^{130}$Te 					&92.26													  & 92.13					        	        & 34.08					           				    					\\
\hline                                         
$^{128}$Te 					&7.71														  &7.28 					      	          & 31.74					                         				\\
\hline
$^{126}$Te 					&0.015													  &0.02 					      	          &18.84 					                          			\\
\hline
$^{125}$Te 			    &0.006											  	  & 0.01					      	          & 7.07					                       					\\
\hline
$^{124}$Te 					&0.0005								 		   	 & $\leq$	0.005				          & 4.74					                       					\\
\hline
\hline
\end{tabular}
\end{table}
\begin{table}[b] 
\centering
\caption{Concentration of the most problematic metallic impurities in enriched metal and in the $^{130}$TeO$_2$ powder used for the $^{130}$TeO$_2$ crystals growth. 
Last column: same values for a  sample of natural TeO$_2$ powder.}
\label{tab:impurities-concentrations}   
\begin{adjustbox}{max width=0.48\textwidth}    
\begin{tabular}{lcccc}
\hline\noalign{\smallskip}
Element    						&$^{130}$Te metal		      				&$^{130}$TeO$_2$ powder 								 &Nat. TeO$_2$ powder 		    	    		\\
													  & [\textit{ppm}] 		 						        &[\textit{ppm}]       				   		  				    			 &[\textit{ppm}]    										     \\
\hline          
Cu												&3.3																		    &$<$0.19					      	            					   	&$<$0.19				           					     					\\
\hline    
Pb							 					&$<$0.017															&$<$0.02 					   										   	        & 0.026					                       			      	\\
\hline				
Al								 					&4.4															  	      &3.4 					   								   	                  &$<$1.9 					                            			\\
\hline
Fe							 			    &0.3												                &$<$0.2				      	                               &1.0		                                    					\\
\hline
Cr							 					&$<$0.09											            &$<$0.09							                              & 0.15				                              					\\
\hline
Ni 												&$<$0.09									              &$<$0.09							                              & 0.09					                            					\\
\hline        
\hline
\end{tabular}
\end{adjustbox}
\end{table}
The  $^{130}$TeO$_2$  crystals were manufactured by Shanghai Institute of Ceramics of the Chinese Academy of Sciences (SICCAS), P.R. China, following basically the same technology as the one applied for the production of CUORE crystals~\cite{Arnaboldi-2010:2999}. Some specific procedures were applied though, in order to reduce the material losses which resulted in a  $^{130}$TeO$_2$ powder synthesis efficiency $\geq$ 80~\%, and a crystal growth efficiency  $\geq$ 90~\%, meaning an \textit{irrecoverable} loss of 28 ~\%\footnote{In an industrial dedicated synthesis and growing procedure these irrecoverable losses could  be reduced.}.
As shown in Table~\ref{tab:impurities-concentrations},  the synthesis of the $^{130}$TeO$_2$   powder acts  as a purification process reducing  most of the metallic impurities to a concentration 
below 1~ppm.

A dedicated furnace and crucible system were built in order to cope with the relatively small amount of  $^{130}$TeO$_2$ available and a \textit{single}-growth cycle was applied instead of the double-growth process used for the production of CUORE crystals. This last point was adopted in order to decrease the amount of losses of the enriched material.

Also a dedicated temperature gradient and growth regime were applied in order to  minimize any mass transfer between the seed (TeO$_2$ crystal with natural Te isotopic concentration) and the growing crystal. Preliminary tests were made using low enriched material as marker in order to make sure that the isotopic concentration of the feeding $^{130}$TeO$_2$ powder remains unchanged in the grown crystal (a detailed description of enriched $^{130}$TeO$_2$ crystal production will be given in a dedicated article). 
One single enriched crystal ingot was finally grown.  Two crystals were produced out of the single ingot  in order to study possible (radioactive) impurities segregation effects during crystal growth. 
The shapes of the two crystals were fixed by the requirement that the crystals be identical with the maximum total mass yield. 
Two  36$\times$38$\times$52 mm$^3$  435 g crystals were cut and processed (shaped, chemical etched and polished) in a dedicated clean room with special precautions aimed at preventing possible radio-contamination of samples. In order to maximize the Cherenkov light output, four of the surfaces were roughly ground while the other two  (the hard faces) were polished to optical standards.

\section{Experimental technique}
Bolometers are very sensitive calorimeters operated at cryogenic temperatures.  These solid-state detectors share with Ge diodes the capability of achieving excellent energy resolution ($\sim$5 keV FWHM from several keV to several MeV) with sizeable active mass devices. Our bolometers apply the calorimetric (source=detector) approach for  the detection of rare decays: the source isotope is part of the active mass of the detector. The latter consists of two elements: a single crystal that plays the role of the calorimetric mass, and a sensor that measures the amount of energy converted into heat in the crystal, converting the phonon signal into an electrical one.

\subsection{Enriched TeO\texorpdfstring{$_2$} \  \ Bolometers}
To operate  a crystal  as a bolometer it must be coupled with a suitable thermometer; in this work we use  3$\times$3$\times$1~mm$^3$ Neutron Transmutation 
Doped (NTD)  Germanium thermistors~\cite{Wang-1990:3761}, thermally coupled to the crystal via nine epoxy  glue spots of $\sim$600$\:\mu$m diameter and $\sim$ 50$\:\mu$m height. The NTD is a resistive device made of semiconducting material, which converts temperature variations into resistance variations. When the thermistor is biased with a constant current, any resistance variation produces a voltage pulse  that constitutes the signal. 
In addition a $\sim$300~k$\Omega$ resistor, made of a heavily doped meander on a 3.5 mm$^3$ Silicon chip, is attached to each crystal and acts as a heater
to stabilize the gain of the bolometer~\cite{Alessandrello-1998:454}. 

\begin{figure}[hbt] 
\centering 
\includegraphics[width=0.4\textwidth]{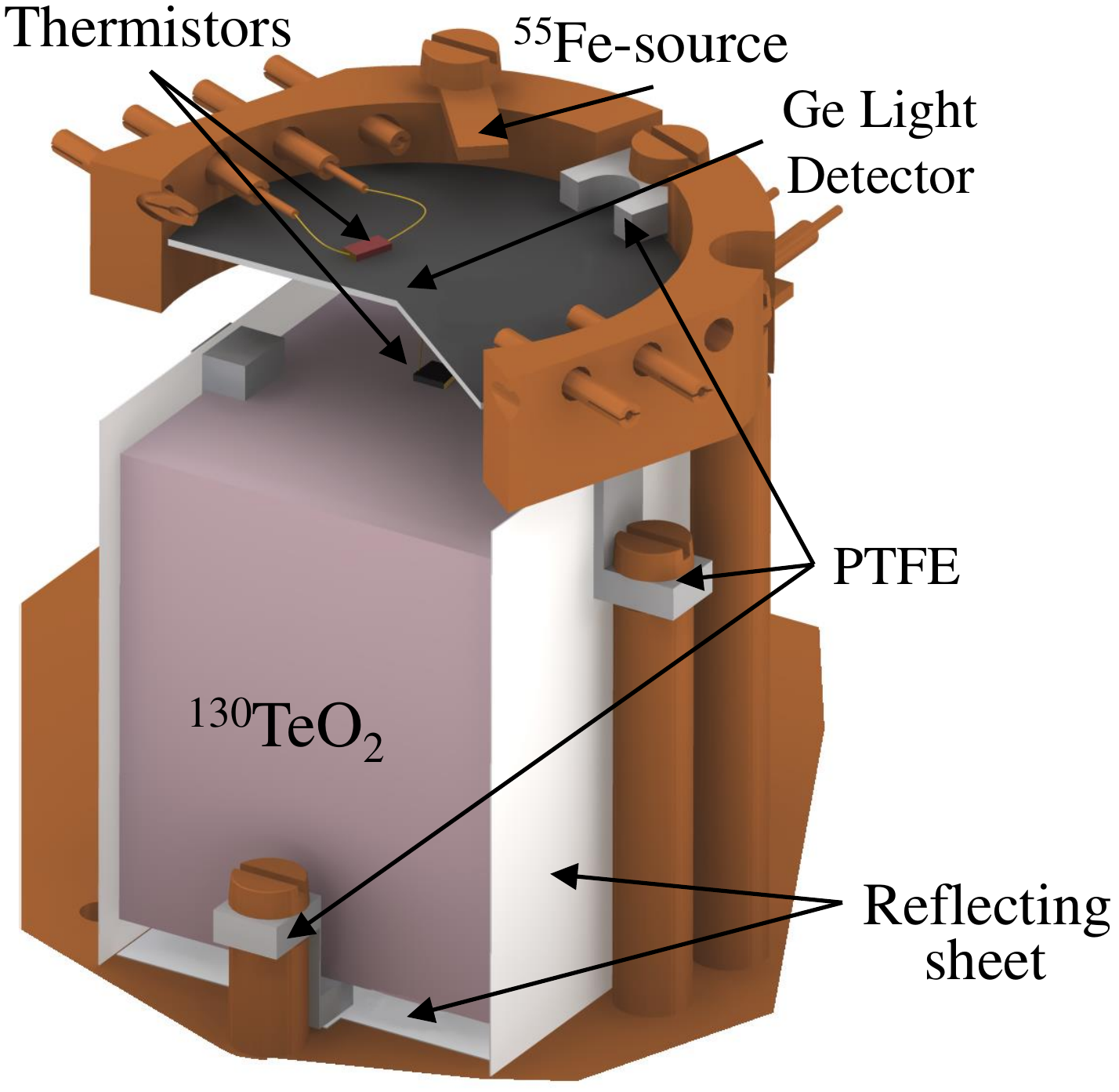}
\caption{Schematic  view of the single module detector. The crystal is surrounded by the reflector sheet in order to enhance the light collection towards the LD, mounted
on the top, facing one of the optical polished surface of the crystal.}
\label{fig:setup}
\end{figure}

The crystal is held by means of four  S-shaped PTFE supports mounted on Cu columns (see Fig.~\ref{fig:setup}). These  Teflon supports ensure that with the down-cooling of the  set-up the crystal is clasped tighter (PTFE thermal contraction is one of the highest among different materials),  to minimize heat-noise generated by  frictions induced by the acoustical vibration of the cryogenic facility. In order to increase the light collection, the crystal is  surrounded laterally and on the bottom part (with no direct thermal contact) by a plastic  reflecting sheet (3M Vikuti$^{TM}$ ESR), while the LD is faced to the top part.
\subsection{Neganov-Luke light detectors}
\label{subsection:neganov-luke}

The  LDs  developed for DBD scintillating bolometers experiments~\cite{Beeman-2013:P07021} generally consist of Ge wafers coupled to NTD Ge thermistors.
The performances of these devices are well satisfactory to read out the scintillation light (few keV) while they  are generally insufficient  to separate $\alpha$ and $\beta$ particles on an 
event-by-event basis in the Region of Interest  (RoI) for 0$\nu$-DBD of $^{130}$Te,  their RMS baseline ($\sim$100~eV) being comparable with the  weak Cherenkov light signal (even if values  $\sim$30--50~eV RMS were recently obtained~\cite{Artusa2016:364}). 

The two LDs used in this work have essentially the same structure and materials, and especially the same temperature sensor, but their  signal-to-noise ratio can be significantly improved by exploiting the Neganov-Luke effect~\cite{Neganov:1981,Luke:1988}. 
This effect is based on the application of an electric field  in the light-absorber volume. The work done by the field on the drifting charges
(generated by the absorption of scintillation light)  is converted into additional heat, which amplifies considerably the thermal signal provided by the NTD Ge thermistor. In our case, the field is generated through a set of concentric Al rings, electrically connected by means of ultrasonic wedge bonding with an alternate pattern. This allows the application of  a  voltage drop (V$^{grid}$) between   adjacent rings and the production of an electric field parallel to the surface. This ring structure enables increasing the collecting field for a given applied voltage and decreasing the charge trapping probability thanks to the short path length of the charges to the electrodes.

\begin{figure}[hbt] 
    \centering
    \includegraphics[width=0.4\textwidth]{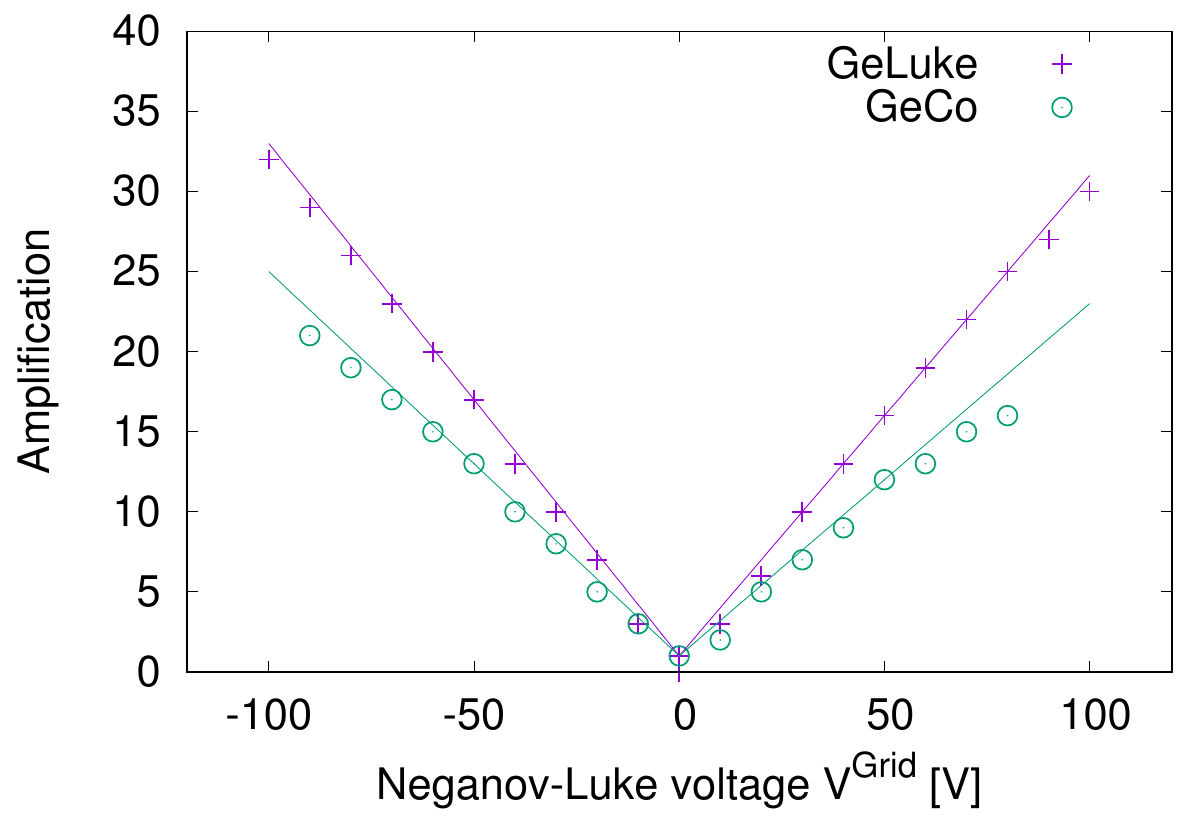}
\caption{Amplification as a function of the applied voltage achieved by Neganov-Luke effect with the GeLuke  and GeCo light detectors. The measurement was performed at 18 mK and light pulses were generated by a 820 nm LED excited with 5 $\mu$s-width voltage square pulses. The GeCo detector develops parasitic currents  above 50 V with a consequent amplification reduction. Linear fits are shown. A slight asymmetry is observed with respect to voltage polarity. }
    \label{fig:NLgain}
\end{figure}

The two Ge LDs (named GeLuke and GeCo in the following) have  a diameter of 44 mm and a thickness of 0.17 mm. They  were previously tested above ground at CSNSM in a dilution refrigerator dedicated to the development of luminescent bolometers~\cite{Mancuso:2014b,Mancuso:2014a}. 
In particular  the amplification induced by the Neganov-Luke effect as a function of the voltage across the electrodes for a given light pulse was measured. The light was guided to the electrode surface of the Ge absorber through an optical fibre by a room temperature LED. The LED wavelength was 820 nm, in the near infrared, with a corresponding absorption length in Germanium of $\sim 0.2$~$\mu$m. An example of the achieved amplification for both LDs is shown in Fig.~\ref{fig:NLgain}. We remark  that - for a fixed  deposited energy -  the amplification for Cherenkov absorption is expected to be lower than the amplification  for LED light absorption, since the Cherenkov photons are distributed in the optical and near-ultraviolet frequency range. 
Therefore, the individual photon energy is higher with respect to LED excitation, decreasing the efficiency in producing electron-hole pairs. In fact, by comparing the LED data in Fig.~\ref{fig:NLgain} with the Cherenkov data in Table~\ref{tab:LD-main-parameters}, a gain reduction by a factor $\sim 1.5$ is observed in both  LDs in the Cherenkov case.

The LDs are energy-calibrated with 5.9~keV and 6.5~keV photons from a $^{55}$Fe X-ray source. This calibration method however works only when the detector is operated at zero bias on the electrodes, since the Neganov-Luke effect broadens and distorts the X-ray peaks~\cite{Willers:2015}. However, it is possible to select a Cherenkov light signal corresponding to a 
well-defined $\gamma$/$\beta$ energy measured in the TeO$_2$ absorber. As explained in Sec.~\ref{sec:result-luke}, the energy associated to this light pulse can be determined at zero Neganov-Luke bias using the X-ray calibration. This energy value can then be used to calibrate the detector response under Neganov-Luke effect and evaluate the amplification for Cherenkov photons. 
\subsection{Experimental set-up}
The detector, sketched  in  Fig.~\ref{fig:setup}, was enclosed in a copper shield. The entire set-up was mounted in the \mbox{CUPID}~R\&D cryostat at Laboratori Nazionali del Gran Sasso. The cryostat consists of a $^3$He/$^4$He  - wet - dilution refrigerator (Oxford TL 200). The detector was cooled down to a temperature of $\sim$12 mK. A complete description of the set-up and the electronics can be found in~\cite{Pirro-2006:672,Arnaboldi-2006:826,Arnaboldi-2004:578}.

The thermistors of the detectors are biased  with a constant voltage through large (few G$\Omega$) load resistors~\cite{Arnaboldi-2002:1808}, resulting in a constant current operation. The resistance variations, generated by the temperature rise, are converted into voltage pulses read across the resistive sensors. The heat and light voltages are then amplified, filtered by a 6-pole Bessel-filter (with a cut-off  frequency of 8~Hz for the $^{130}$TeO$_2$ crystals and 120~Hz  for the LDs)  and finally fed into a NI PXI-6284 18-bits ADC.

The sampling rate of the ADC is 2~kHz  for the $^{130}$TeO$_2$ crystals and 4 kHz for the LDs. All triggers are software generated: when a trigger fires, for the main bolometer and 
the LD, waveforms 2.5~s and 0.25~s long are recorded. Moreover, when the trigger of the $^{130}$TeO$_2$ crystal fires, the corresponding waveform from its LD is always recorded, irrespectively of its trigger. The amplitude and the shape of the voltage pulse is then determined by the off-line analysis. The pulse amplitude of the thermal signals are estimated by means of the Optimum Filter (OF)  technique~\cite{Gatti-1986:1,Alduino-2016:045503}, that maximizes the signal-to-noise ratio in a way that improves the energy resolution and lowers the threshold of the detector. The amplitude of the  light signals, however, is evaluated from the filtered waveforms at a fixed time delay with respect to the $^{130}$TeO$_2$ bolometer, as described in detail in~\cite{Piperno-2001:10005}.\newline
The amplitude of the acquired heat spectrum is energy-calibrated using several $\gamma$-ray peaks arising from the calibration with $^{228}$Th and, if present, known high-energy 
$\alpha$-lines. The energy scale is linearised  with a second order polynomial function in $log (V)$ with zero intercept,  where V is the heat
pulse amplitude.
The LDs, on the contrary, are calibrated thanks to the  5.9  and 6.5 keV lines  produced by the permanent $^{55}$Fe X-ray sources faced  to the detectors (see Fig.~\ref{fig:setup}).
\newline
The data sets  analysed  here consist of a first background campaign, in which only the two $^{130}$TeO$_2$ crystals were acquired,  and several calibration runs to optimize 
the $\alpha$ vs $\beta/\gamma$ rejection power  by changing the working  conditions of the LDs and cryostat operation.  
\section{Data analysis and results}
\label{sec:results}
\subsection{\texorpdfstring{$^{130}$TeO$_2$}  \ { \ performances}}
\begin{table}[b] 
\centering
\caption{Main parameters of the $^{130}$TeO$_2$ bolometers.  The third column represents the theoretical resolution given by the Optimum Filter, while the last one
represents the absolute signal read out across the thermistor. These values  are consistent  and comparable with the CUORE-0 natural TeO$_2$ crystals~\cite{Alduino-2016:P07009}.}
\label{tab:main_parameters}       
\begin{tabular}{lcccc}
\hline\noalign{\smallskip}
																				&R$_{work}$     			&$\tau_	{decay}	$		 &baseline noise 							 &Signal         			       \\
																				&[M$\Omega$]   			& [ms] 		 						     &[keV FWHM]       				   &[$\mu$V/MeV]         \\
\hline
$^{130}$TeO$_2$-1     			&286		   								&200														   & 3.5                      &135                        \\
\hline
$^{130}$TeO$_2$-2   		    &154		    							&145														   &4.2            	    	   &95                     	   \\
\hline		
\end{tabular}
\end{table}

\begin{figure}[hbt] 
\centering 
\includegraphics[width=0.45\textwidth]{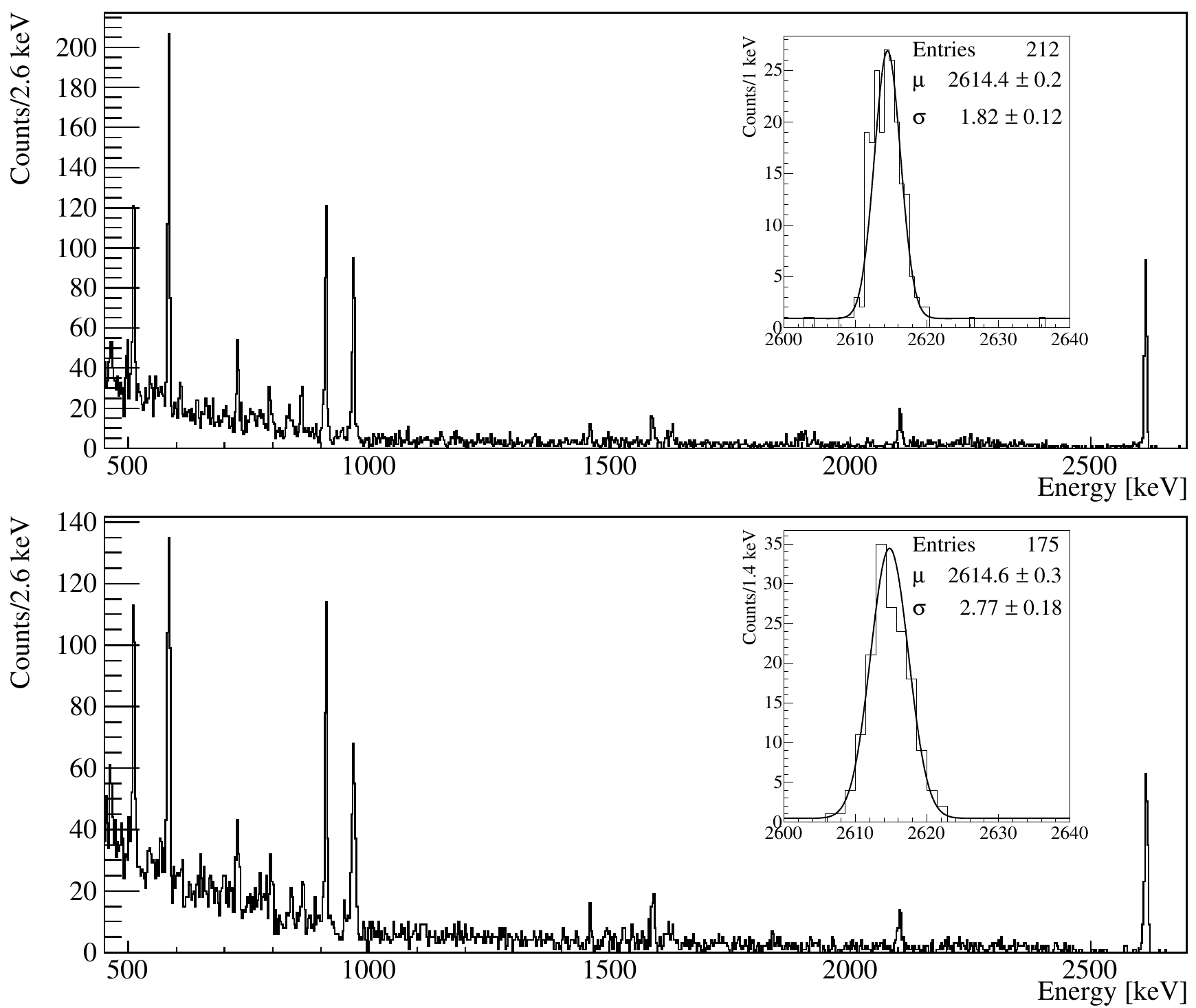}
\caption{$^{228}$Th calibration spectra collected over 64 hours. Top $^{130}$TeO$_2$-1; bottom $^{130}$TeO$_2$-2. The 2615 keV $\gamma$ peak of $^{208}$Tl is highlighted in the inset.}
\label{fig:calibration}
\end{figure}

The crystals were  operated at a temperature of $\sim$13 mK. The most important parameters of the bolometers are shown in Table~\ref{tab:main_parameters}.
In Fig.~\ref{fig:calibration} we present the calibration spectra of the two $^{130}$TeO$_2$ crystals: the energy resolution close to the 0$\nu$-DBD RoI 
of $^{130}$Te, evaluated with the 2615 keV line of $^{208}$Tl,  is 4.3 and 6.5 keV FWHM, respectively. These  values  are fully compatible 
with the ones obtained in the CUORE-0 experiment~\cite{Artusa-2014:2956}, performed with natural  TeO$_2$ crystals.

\begin{figure}[hbt] 
\centering 
\includegraphics[width=0.45\textwidth]{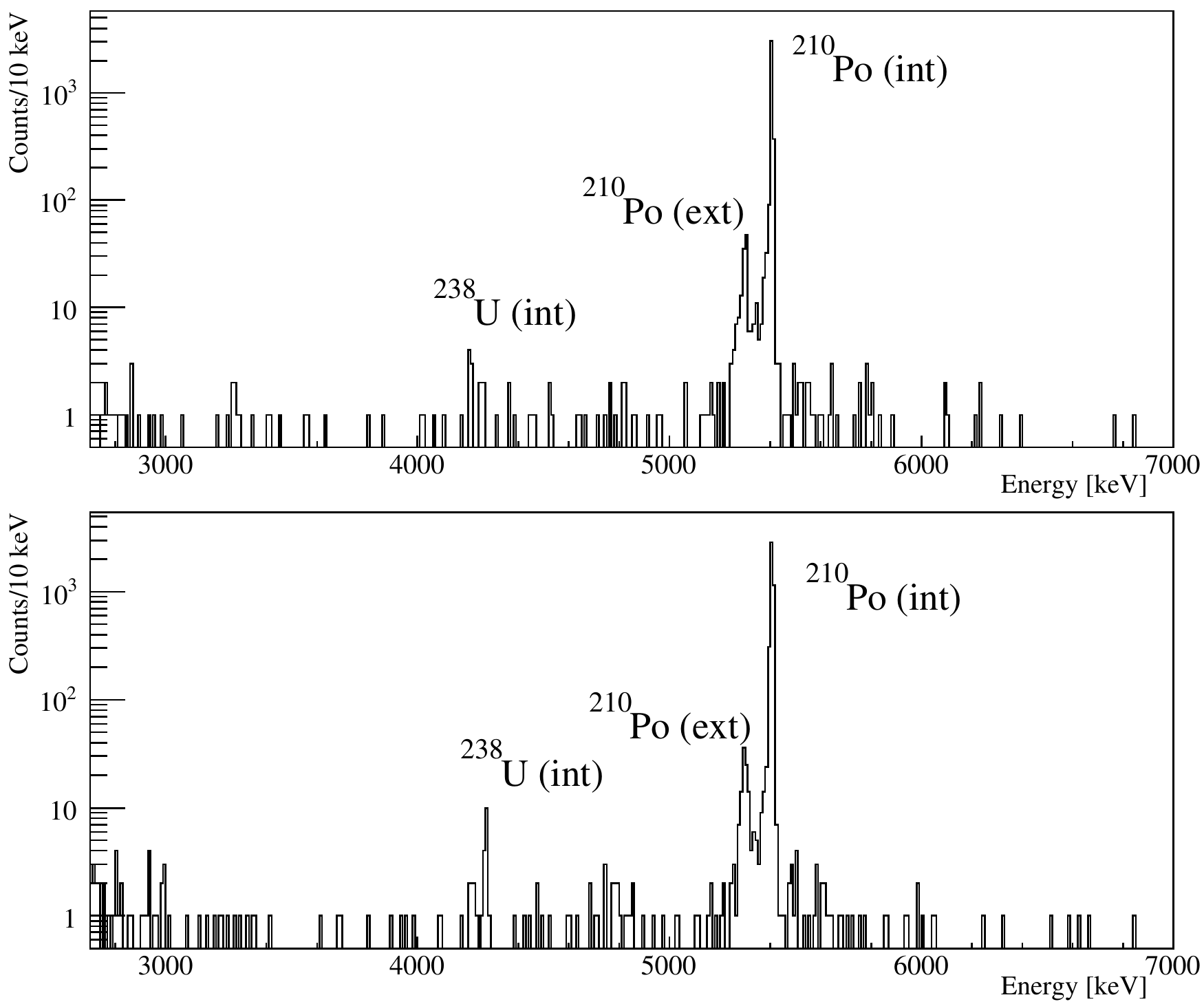}
\caption{Alpha energy region. The total statistics of $^{130}$TeO$_2$-1 corresponds to 434.3 h of background + 229.5 h of calibration (total 663.8 h), while the 
one of   $^{130}$TeO$_2$-2 consists of 337.1 h of background + 168.7 h of calibration, totalling 505.8 h of  live time. The contamination in $^{210}$Po shows also a surface 
contribution (due to contamination of the surface of the crystal and/or of the surface facing it) due to the escape of the nuclear recoil of $^{206}$Pb carrying out 103 keV.}
\label{fig:alpha-background}
\end{figure}

Background measurements were performed  to assess the internal contaminations of the enriched crystals. This,  together with the energy resolution, was the most important issue for this work, since previous experience with enriched  TeO$_2$  crystals did not give 
satisfactory results  for either energy resolution~\cite{Pirro-2000:71} or for  internal background~\cite{Alessandrello-2000:13}.
Since, after the run was ended, the analysis gave only limits on the most worrisome internal contaminants, we decided to use also the  
$^{228}$Th calibration runs  in order to increase  our sensitivity in the high energy $\alpha$-region. The final spectra of the $\alpha$-region of the two  $^{130}$TeO$_2$ crystals is presented in  Fig.~\ref{fig:alpha-background}.
The only peaks appearing in the spectrum correspond to a clear signal  due to $^{210}$Po and to a tiny contamination  of   $^{238}$U.

$^{210}$Po is a very well known contamination of TeO$_2$ crystals and is present (both internally and on the surface)  in all the TeO$_2$ crystals produced for CUORE~\cite{Alessandria-2012:839}. 
Due to the relatively short decay time, this isotope does not represent a problem for  a DBD search. $^{238}$U, on the contrary, was not observed in the CUORE natural crystals, 
with a detection  limit of 5$\times$10$^{-14}$ g/g (corresponding to 0.6 $\mu$Bq/kg).
 
No other contaminations are visible  in the spectra. The results are presented in Table~\ref{tab:alpha-activity}. 
To obtain the limits, we defined the signal as the number of events falling in the energy region within $\pm3\sigma$   of the  Q$^\alpha_{value}$ and the background as the average number of events 
falling in the 3$\sigma$ side-bands of this interval, being $\sigma$ the energy resolution of the detector. Following the Feldman-Cousins approach, we computed the 90~\% C.L. upper limit on the number of events and we inferred the upper limit on the 
activity. Finally, the limits for $^{226}$Ra and $^{228}$Th are further improved by exploiting the lack of evidence of $\alpha$ delayed-coincidences of their daughter-nuclei.  Internal $^{226}$Ra can be evaluated 
by exploiting the  unique time and energy stamp given by the decay of   $^{222}$Rn to $^{218}$Po  followed by   the 46.1 min delayed high energy decay of  $^{214}$Bi and $^{214}$Po  (\textit{Bi-Po} events).  The decay of  $^{224}$Ra to $^{220}$Rn and $^{216}$Po, furthermore,  gives a second   unique stamp for the evaluation of $^{228}$Th internal contaminations.
This search, performed within a 4 $\tau_{decay}$ time interval  - with respect to each specific decay -,  did not give evidences of any events.
\begin{table}[!b]
\centering
\caption{Activity of trace contaminations   belonging to $^{232}$Th  and $^{238}$U chains for the two crystals. The total collected statistic is 663.8  hours for 
$^{130}$TeO$_2$-1	and 505.8 hours for $^{130}$TeO$_2$-2. Limits at 90~\% C.L. See text for more details.}
\begin{tabular}{lccc}
\hline
Chain           								   &Nuclide       		              &$^{130}$TeO$_2$-1						  &$^{130}$TeO$_2$-2			\\
																		 &                                   &[$\mu$Bq/kg] 								  			&[$\mu$Bq/kg] 								\\
\hline
$^{232}$Th   									 &$^{232}$Th             			&$<$4.3 																&$<$4.8													\\
\hline
																		&$^{228}$Th             			&$<$2.3																	&$<$3.1 												\\
\hline
\hline
$^{238}$U  									  &$^{238}$U             		    &7.7 $\pm $ 2.7 									    &15.1 $\pm$ 4.4							\\
\hline
																		&$^{234}$U              		   &$<$6.3 											   				&$<$5														\\
\hline
																		&$^{230}$Th           	      &$<$5.7 									 								&$<$3.8													\\
\hline
																		&$^{226}$Ra              			&$<$2.3 																		&$<$3.1												\\
\hline
																		&$^{210}$Po 		              & 3795 $\pm$ 60							&6076 $\pm$ 88						\\
\hline
\end{tabular}
\label{tab:alpha-activity}
\end{table}
\subsection{Light detector performance}
\label{sec:result-luke}
Most of the calibrations with the $^{228}$Th  source were performed to study and optimize the Neganov-Luke amplified LDs. The first step, the energy calibration of the 
light signal, was a 45 h long measurement with the   $^{228}$Th  source, with V$^{grid}$ set to zero. Despite the fact that the RMS baseline at  V$^{grid}$ =0 is of the same order of the signal,  the high statistical value of the data and the use of the light-synchronization~\cite{Piperno-2001:10005}   allowed the mean energy of the Cherenkov light from the 2615 keV  $\gamma$-rays
to be accurately converted to an energy value using the $^{55}$Fe X-ray calibration.
The obtained values for the absolute Cherenkov light signals  are 153$\pm$4 eV and 160$\pm$5 eV for   $^{130}$TeO$_2$-1		and		$^{130}$TeO$_2$-2, respectively.
These values are in good agreement with the ones obtained with natural TeO$_2$ crystals of similar size ~\cite{Casali-2015:12,Schaeffner-2015}.
This monochromatic light signal, independent from any parameter, is then used to calibrate the gain of the detectors once a V$^{grid}$ is applied and the   $^{55}$Fe peaks are
no longer clearly identifiable  (due to the presence of  V$^{grid}$).
Several calibrations were performed with different values of  V$^{grid}$ in order to evaluate the  best signal-to-noise ratio of the LDs. The best compromise between gain and noise
was found at  V$^{grid}$=25 V for GeLuke and V$^{grid}$=55 V for GeCo. The main parameters of the LDs are shown in Table~\ref{tab:LD-main-parameters}. 
\begin{table}
\centering
\caption{Main parameters of the two LDs bolometers. The gain is evaluated as the amplitude of the Cherenkov light signal induced by the absorption of a 2615 keV $\gamma$ quanta in the  $^{130}$TeO$_2$  at V$^{grid}$=25 V (GeLuke) and V$^{grid}$=55 V (GeCo), divided by the corresponding value obtained at V$^{grid}$=0 V. The last column, instead, represents the baseline noise obtained by applying the  V$^{grid}$ voltage. The fact that RMS/RMS$^{V^{grid}}<$ Gain clearly indicates the presence of excess  noise induced by injection of the bias to the grid.}
\begin{adjustbox}{max width=0.48\textwidth}
 \begin{tabular}{lccccc}
\hline\noalign{\smallskip} 
																				&R$_{work}$     				 &Signal										 &RMS  					&Gain 															&RMS      					    								 \\			   
																				&[M$\Omega$]   	   &[$\mu$V/MeV]             &[eV]       			&(V$^{grid}$)				        			&[eV]    (V$^{grid}$) 					    	\\
\hline
GeLuke											  			&2.4		   									 &   570       		 	      		&166             &5.8  (25 V)     									& 35 (25 V)	                        		\\
\hline	
GeCo											  		    &2.3	    								   & 1320        	  	  	    	 &87             & 8.9  (55 V) 	   		  		 	 			& 25	(55 V)  										   			\\
\hline		
\end{tabular}
\end{adjustbox}
\label{tab:LD-main-parameters}     
\end{table}
\begin{figure*}[ht]
\begin{center}
\includegraphics[width=0.8\linewidth]{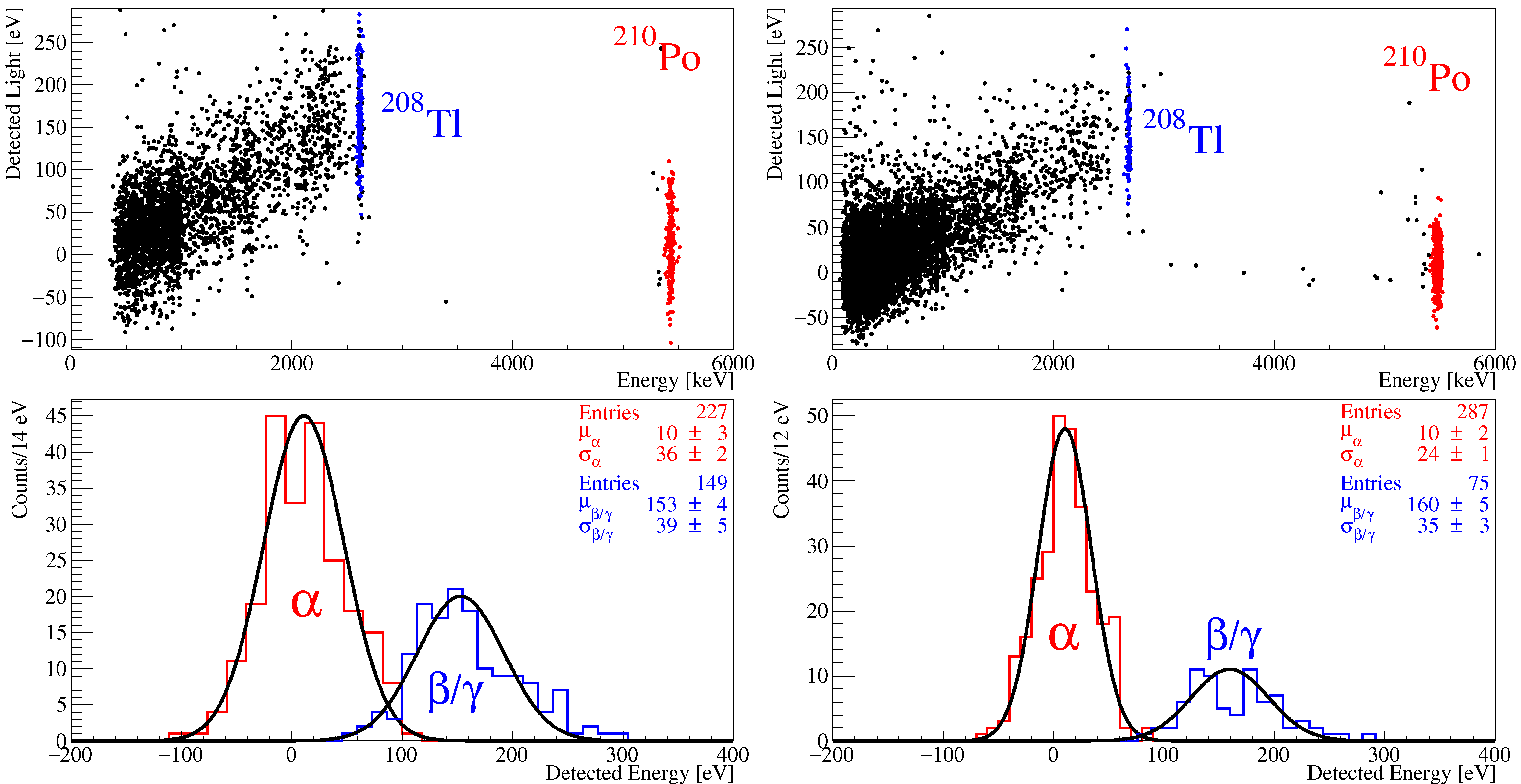}
\end{center}
\caption{Top: Light vs Energy scatter plots obtained for $^{130}$TeO$_2$-1 + GeLuke (left) and $^{130}$TeO$_2$-2 +  GeCo (right)  with 30 and 36  hours  $^{228}$Th  calibration, respectively. Bottom:
corresponding distributions of the light signals due to $^{208}$Tl-2615 keV $\gamma$s  and   $^{210}$Po-5407 keV $\alpha$s  interactions for the same detectors. Each distribution, chosen with a cut on the heat-energy signal corresponding to an interval of $\pm$ 2$\sigma$ around the central value of each peak, is fitted with a simple Gaussian function.  See text for more details.}
\label{fig:scatter-plots}
\end{figure*}
\subsection{Particle identification}
\label{sec:particle-ID}
In Fig.~\ref{fig:scatter-plots} (top)  we show  the  Light  vs Energy scatter plot obtained with the two enriched crystals.
There is a clear separation between the $\beta/\gamma$ induced events and the $\alpha$-events (mainly due to $^{210}$Po). 
In order to  evaluate the $\beta/\gamma$ vs $\alpha$  Discrimination Power (DP), we select the light signals 
belonging to the  2615 keV  $^{208}$Tl $\gamma$-line  and   the 5407 keV $^{210}$Po $\alpha$-line, taken as signal and background, respectively. In the bottom part of Fig.~\ref{fig:scatter-plots} the two light distributions are fitted with a Gaussian function.
The  DP between the two distributions can be quantified as  the difference between the average values of the two distributions  normalized to the square root of the quadratic sum of  their  widths: 
\begin{equation} 
DP = \frac{|\mu_{\alpha}-\mu_{\gamma/\beta}|}{\sqrt{\sigma^{2}_{\alpha} + \sigma^{2}_{\gamma/\beta}}}.
\label{eq:DP}
\end{equation} 
\begin{figure}[t] 
\centering 
\includegraphics[width=0.5\textwidth]{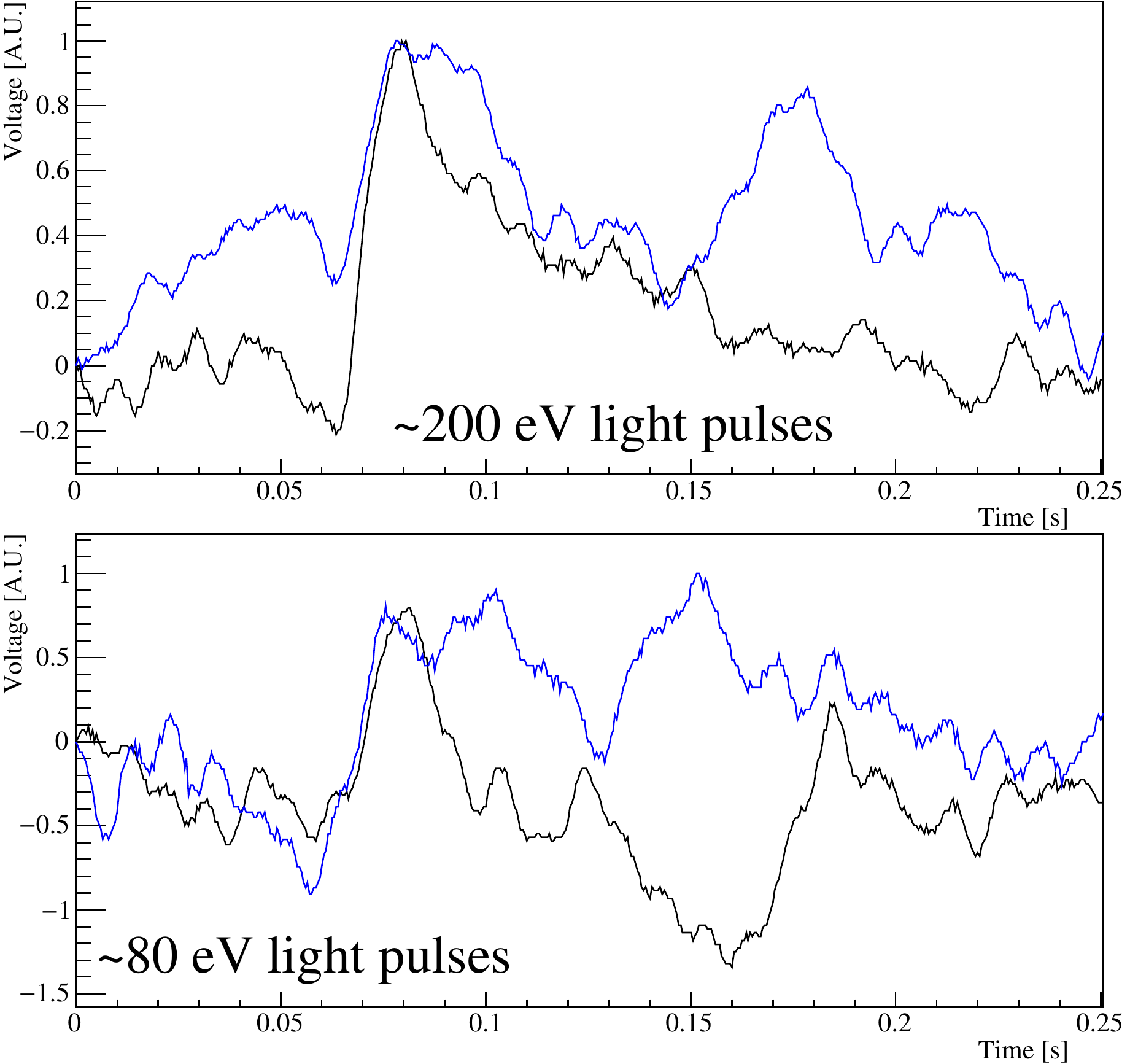}
\caption{Typical Cherenkov light signals read out by the GeCo LD. Top: signals corresponding to $\approx$200 eV. Bottom: signals corresponding to $\approx$80 eV. The different colours 
simply distinguish between a pulse that can be eye-recognized (black) and a pulse that is "confused" within micro-phonic noise (blue). All signals are aligned at the maximum reconstructed amplitude of the pulse, 80 ms.}
\label{fig:Cherenkov-Pulses}     
\end{figure}
Using the fit values shown  in Fig.~\ref{fig:scatter-plots}, we get a DP = 2.65 for   $^{130}$TeO$_2$-1 (GeLuke) 	and DP = 3.5 for the  $^{130}$TeO$_2$-2 (GeCo), respectively. 
Choosing an acceptance level of 95~\% for  the e/$\gamma$ signal, we get  $\alpha$  rejection factors of  98.21~\% and  99.99~\%, respectively.

For the sake of completeness, two important remarks should be made. The first regards the fact that to correctly evaluate the efficiency of a particle ID,  the two classes of events should have the same energy, while in this case  the light-signal distributions we compared  were  generated by   2615 keV $\gamma$s  and 5407 keV $\alpha$s.

However,  since the $\alpha$-particles do not emit photons (the mean energy of the two light distributions - as shown in Fig.~\ref{fig:scatter-plots} - is very close to  zero\footnote{In fact a small deviation from zero could be induced by a very small cross talk between light and heat channels  or by a very weak scintillation light emitted by the TeO$_2$ crystal in addition to Cherenkov light, 
as the results in Ref.~\cite{Coron-2004:159} seem to suggest. In any case, this behaviour  slightly underestimates  the real DP at lower energies.}), this point does not hold.
This can be also deduced by the fact that  the width of the $\alpha$ light signal distribution of the two detectors (see Fig.~\ref{fig:scatter-plots}) is fully compatible with the RMS noise of 
the LDs, as given in Table~\ref{tab:LD-main-parameters}. 

The second point is that the  Q$_{\beta\beta}$  value of    $^{130}$Te is at 2528 keV, slightly less than 2615 keV.
Since the Cherenkov light, at first level,  is proportional to the energy, this would correspond to a  slightly lower DP in the RoI. However, simulations
show~\cite{Casali-2016-arXiv}   that the Cherenkov light generated by  two electrons (i.e. the 0$\nu$-DBD signal) sharing 2528 keV is statistically larger with respect to the one 
generated by a $\gamma$ of the same energy. In fact this last effect overtakes the first one so that, actually,  the evaluated DP  at 2615 keV slightly underestimates the one 
at Q$_{\beta\beta}$ for  the 0$\nu$-DBD detection.

Finally,  to have a more clear picture of the extremely small amplitude  of the acquired Cherenkov light signals, in Fig.~\ref{fig:Cherenkov-Pulses} we plot four  pulses corresponding to two different energies.
These pulses come from  the tails (left and right) of the $^{208}$Tl $\gamma$ light distribution  of Fig.~\ref{fig:scatter-plots} (bottom right). Thanks to the Optimum Filter technique~\cite{Gatti-1986:1,Alduino-2016:045503}  and, especially, to the trigger synchronization~\cite{Piperno-2001:10005}  signals that cannot be  auto-triggered, give rise to a very powerful
$\alpha$ vs $\beta/\gamma$ event-by-event identification.
\section{Conclusions}
In this work we clearly demonstrated for the first time the possibility to operate large enriched  $^{130}$TeO$_2$ crystals to search for DBD with the bolometric technique. The obtained energy resolution is compatible with those obtained with natural crystals that will soon be operated in the CUORE experiment. The internal radioactive contaminations show  only a very small contribution 
from  $^{238}$U. The corresponding secular equilibrium of the decay chain is broken and, finally,  the internal activity of the most problematic nuclei for 0$\nu$-DBD, both  $^{226}$Ra and $^{228}$Th are evaluated   as $<$3.1 (2.3)  $\mu$Bq/kg, respectively. Both the enriched crystals have the $^{226}$Ra contamination limit  that is within the specification 
of the CUORE crystals ($<$3.7 $\mu$Bq/kg), while with respect to $^{228}$Th, the obtained limits are very close to the specification ($<$1.2 $\mu$Bq/kg). These values  represent a very good starting point 
even if  strong efforts are needed for a further reduction of the internal contaminations.
Furthermore we finally demonstrated that the weak Cherenkov signal can be read with \textit{standard} thermistor-based bolometers  and used to discriminate the $\alpha$ induced background with a rejection  factor more than 100, the goal of the CUPID project.
 \section{Acknowledgements}
This work was supported by the LUCIFER experiment, funded by the European Research Council under the Seventh Framework Programme (FP7 2007-2013) ERC Grant agreement 247115,
by the Italian Ministry of Research under the PRIN 2010ZXAZK9 2010-2011 grant, by the US National Science Foundation (Grant n.0605119 and n.1307204), by  the National Natural Science Foundation of China, project n. 51302287 and n. 61405229, by the U. S. Department of Energy National Nuclear Security Administration under Award No. DE-NA0000979 and was performed under the  auspices of US Department of Energy by LLNL under contract DE-AC52-07NA27344.
The development of the LDs  was supported by the LUMINEU program, receiving funds from the Agence Nationale de la Recherche (France).

We wish to express our gratitude to LNGS and, in particular, to the mechanical workshop in the person of  E. Tatananni, A. Rotilio, A. Corsi, and B. Romualdi 
for continuous,  constructive and tireless  help in the overall set-up construction. We are  grateful 
also to  M. Guetti for his invaluable support in the cryostat facility maintenance.
\section*{References}
\bibliography{Enriched-Te}
\end{document}